\documentclass[useAMS,usenatbib,epsf]{mn2e}
\usepackage{graphics}
\title[Near-IR polarimetry and modelling of \textit{IRAS}
19306+1407.]{Near-infrared polarimetry and modelling of the dusty
young PN \textit{IRAS} 19306+1407.}
\author[K. T. E. Lowe \& T. M. Gledhill]{K.~T.~E.~Lowe\thanks{E-mail:
{\tt klowe@star.herts.ac.uk}} and T.~M.~Gledhill\thanks{E-mail:
{\tt t.gledhill@star.herts.ac.uk}} \\
Centre for Astrophysics Research,  S.T.R.I.,  University of
Hertfordshire,
College Lane,  Hatfield, Hertfordshire AL10 9AB, UK}
\date{Released 2006 - accepted 29th September 2006 MNRAS}

\pagerange{\pageref{firstpage}--\pageref{lastpage}} \pubyear{2006}

\def\gsim{\mathrel{\rlap{\lower4pt\hbox{\hskip1pt$\sim$}}
    \raise1pt\hbox{$<$}}}

\begin{document}
\input{epsf}
\maketitle

\begin{abstract}
We present near-infrared polarimetric images of the dusty circumstellar envelope (CSE) of {\it IRAS} 19306+1407, acquired at the United Kingdom Infrared Telescope (UKIRT) using the UKIRT 1-5~$\umu$m Imager Spectrometer (UIST) in conjunction with the half-waveplate module IRPOL2.  We present additional 450 and 850~$\umu$m photometry obtained with the Sub-mm Common User Bolometer Array (SCUBA) at the James Clerk Maxwell Telescope (JCMT), as well as archived {\it Hubble Space Telescope} ({\it HST}) {\it F606W}- and {\it F814W}-filter images.  The CSE structure in polarized flux at {\it J-} and {\it K-}bands shows an elongation NNE-SSW with two bright scattering shoulders NW-SE. These features are not perpendicular to each other and could signify a recent `twist' in the outflow axis.  We model the CSE using an axisymmetric light scattering (ALS) code to investigate the polarization produced by the CSE, and an axisymmetric radiation transport (DART) code to fit the SED. A good fit was achieved with the ALS and DART models using silicate grains, 0.1-0.4~$\umu$m with a power-law size distribution of $a^{-3.5}$, and an axisymmetric shell geometry with an equator-to-pole contrast of 7:1.  The spectral type of the central star is determined to be B1{\sc I} supporting previous suggestions that the object is an early PN.  We have constrained the CSE and interstellar extinction as 2.0 and 4.2~mag respectively, and have estimated a distance of 2.7~kpc.  At this distance the stellar luminosity is $\sim$4500~L$_{\sun}$ and the mass of the CSE $\sim$0.2~M$_{\sun}$.  We also determine that the mass loss lasted $\sim$5300~yrs with a mass-loss rate of $\sim$3.4$\times$10$^{-5}$~M$_{\sun}$~yr$^{-1}$. 
\end{abstract}

\begin{keywords}
stars: AGB and post-AGB -- stars: circumstellar matter -- infrared: stars --
stars: individual (\textit{IRAS} 19306+1407) -- stars: mass loss --
techniques: polarimetric.
\end{keywords}

\section{Introduction}

Post-asymptotic giant branch (post-AGB) stars are luminous
(10$^3$-10$^4$~L$_{\sun}$) evolved stars with initial masses in the
range 0.8-8~M$_{\sun}$ \citep*[see][for a general review]{vw03}. At
the end of the AGB phase, mass-loss rates can peak at over $10^{-4}$
M$_{\sun}$~yr$^{-1}$ before dropping dramatically, as the star enters
its post-AGB evolution \citep*[e.g.][]{s83}, creating detached envelopes
of gas and dust. These dusty circumstellar envelopes (CSEs) are then
visible at optical and near-infrared wavelengths as proto-planetary
nebulae (PPN; \citealt*{k93}). A seemingly ubiquitous feature of PPN is
their lack of spherical symmetry, with many having a bipolar or
point-symmetric structure. Notable and well-studied examples are the
Egg Nebula (AFGL 2688; \citealt{sah98}) and the Red Rectangle (AFGL 915;
\citealt{coh04}). Optical and near-infrared surveys of PPN have shown
that in all cases where a CSE is detected then it appears
asymmetric in some way (e.g. \citealt*{ueta00};
\citealt{gchy01}). Possible mechanisms for the shaping of PPN usually
involve interaction of the mass-losing star with a binary companion,
and have been reviewed by \citet*{bal02}.


Imaging polarimetry is a differential imaging technique, which is
well-suited to the study of CSEs surrounding post-AGB stars. The
technique discriminates between the faint but polarized scattered
light from the PPN and any bright unpolarized emission from the
central star. This enables the imaging of circumstellar material that
would normally be lost under the wings of the stellar point spread
function (PSF), thereby obtaining information on the dust distribution
close to the central source. Imaging polarimetric surveys of post-AGB
stars using the UK Infrared Telescope have detected scattered light
from PPN around 34 stars, and all of these PPN were found to be
axisymmetric in some way (\citealt{gchy01}; \citealt*{g05}). Higher spatial
resolution polarimetry using the Hubble Space Telescope ({\it HST}) has
enabled more detailed studies of the morphology of
PPN, as well as providing constraints on dust grain properties in
these systems, and has revealed point-symmetries, jets and multi-lobed
structures \citep*[e.g.][]{umm05,shk03}.

In this paper, we examine \textit{IRAS} 19306+1407 (GLMP 923), which
has \textit{IRAS} colours typical of a post-AGB star with a cold CSE
\citep{olflhhs93}. Radio and millimetre surveys for molecular emission
have failed to detect OH or H$_2$O masers \citep*{l89} or CO emission
\citep{alk86,lfom91}. However, the object shows a number of dust
spectral features. \citet*{hvk00} present \textit{ISO} spectroscopy
showing emission features at 6.3, 7.8 and 10.7 ~$\umu$m, with a
``probable'' feature at 3.3 ~$\umu$m, and compare these features to
the unidentified infrared (UIR) bands at 3.3, 6.2 and 7.7 ~$\umu$m,
commonly attributed to polycyclic aromatic hydrocarbon (PAH) molecules
\citep*{atb89}. Given that the mid-infrared spectral features are
similar to those seen in hot carbon-rich PN, \citet{hvk00} suggest
that the object is a young PN. A further analysis of the {\it ISO}
data by \citet{hkpw04} confirms the presence of UIR features at 3.3,
6.2, and 7.7~$\umu$m, with the addition of the 8.6 and 11.2~$\umu$m
features. These authors also mention the presence of silicate emission
at 11, 19 and 23~$\umu$m, raising the possibility that \textit{IRAS}
19306+1407 may have a mixed CSE chemistry.

Optical spectroscopy shows a broadened H$\alpha$ emission line with
line width of $\sim$2300~km~s$^{-1}$ indicating a fast outflow
\citep*{sc04}, as well as H$\beta$ and [N{\small II}] emission,
leading \citet*{kh05} to suggest a spectral type of approximately B0
for the star. A number of H$_2$ emission lines are seen in the
$K$-band, with line ratios suggesting a mix of radiative and shock
excitation \citep*{kh05}. Imaging through a narrow-band H$_2$ filter,
centred on the 2.122~$\umu$m line, shows that the H$_2$ emission has a
ring-like structure with evidence for bipolar lobes extending
perpendicular to the ring \citep*{vhk04}.


We present the first near-infrared polarimetric images of the dusty
CSE of {\it IRAS} 19306+1407, showing the structure of the envelope in
scattered light. We also present new sub-millimetre photometry and
archived {\it HST} images. The observations are interpreted using
2-dimensional (axisymmetric) light scattering and radiation transport
models.


\section{Observations and Results}

\begin{table}
\caption{Summary of photometry for \textit{IRAS}
19306+1407 for \textit{HST} (using Vega zero points), UKIRT and SCUBA observations, including
integration time (Int.) and the
extent (Size) of the semi-major and minor axes of the aperture used in
photometry.
The PA angle of photometry aperture is equal to 18\degr (E of N).}
\label{table1}
\begin{tabular*}{\columnwidth}{@{}lccc@{\extracolsep{2pt}}c}
\hline
Band       & Magnitude        & Flux            & Int.  & Size \\
           &                  & (mJy)           & (s)   & (arcsec $\times$ arcsec) \\
\hline
{\it F606W}$^{\rmn{a}}$  & $13.81 \pm 0.03$ & $9.5 \pm 0.3 $            & 300    & $3.2 \times 2.0$\\
{\it F814W}$^{\rmn{b}}$  & $12.45 \pm 0.02$ & $26.1 \pm 0.5 $            & 50     & $3.2 \times 1.9$\\
{\it J}$^{\rmn{c}}$     & $11.18 \pm 0.04$ & $51.5 \pm 0.8$  & 237.6  & $3.9 \times 2.4$ \\
{\it K}$^{\rmn{d}}$     & $10.29 \pm 0.12$ & $48.4 \pm 2.2$   & 237.6  & $3.9 \times 2.4$\\
{\it 450W}$^{\rmn{e}}$   & -              &   $49.9 \pm 38.7$ & 1334$^\dag$ & - \\
{\it 850W}$^{\rmn{f}}$   & -              &   $14.1 \pm 3.7$  & 1334$^\dag$ & - \\
\hline
\end{tabular*}
{\footnotesize Notes: central wavelengths at $^{\rmn{a}}0.5888\umu$m (Broad {\it V}),
$^{\rmn{b}}0.8115\umu$m (Johnson {\it I}),
$^{\rmn{c}}1.25\umu$m,
$^{\rmn{d}}2.2\umu$m,
$^{\rmn{e}}450\umu$m and
$^{\rmn{f}}850\umu$m;
and $^\dag$inclusive of observational overheads.}
\end{table}

\subsection{Imaging polarimetry observations and results}

\begin{centering}
\begin{figure*}
\vspace{15cm}
\caption{{\bf High resolution images are available at http://star-www.herts.ac.uk/$\sim$klowe/}.
The \textit{J-} and \textit{K-}band observations are
displayed at the top and bottom of the figure respectively.  These
images have been scaled logarithmically.  The total intensity (I) is
displayed in sub-figures (a) and (c) with overlaid polarization
vectors (pol).
Sub-figures (a) and (c) are scaled between 20 and 13
mag~arcsec$^{-2}$.  The lowest outer contour levels are 19 and 18
mag~arcsec$^{-2}$ and separated by 1~mag~arcsec$^{-2}$ for (a) and (c)
respectively.  The polarized flux (IP) images (b) and (d) are scaled
between 20 to 16 mag~arcsec$^{-2}$ and 19 to 16 mag~arcsec$^{-2}$
respectively.  The lowest outer contours are 19 (b) and 18 (d)
mag~arcsec$^{-2}$ and separated by 0.5 mag~arcsec$^{-2}$.
}
\label{fig1}
\end{figure*}
\end{centering}

Polarimetric imaging at {\it J}- and {\it K}-band of
\textit{IRAS} 19306+1407 was obtained at the 3.8-m United Kingdom
Infrared Telescope (UKIRT) on Mauna Kea, Hawai'i, using the UKIRT 1-5
micron imager spectrometer (UIST) in conjunction with the infrared
half-waveplate module (IRPOL2). A pixel scale of 0.12 arcsec was used
and observations were made on 2003 June 8 with an average seeing of
0.5 arcsec. The total integration time for each filter was 237.6
seconds, comprising 24 exposures of 9.9 seconds each (see
Table~\ref{table1}). Linear polarimetry was obtained by observing at
four half-waveplate angles of 0$\degr$, 22.5$\degr$, 45$\degr$ and
67.5$\degr$. The data reduction was carried out using
\textsc{starlink}\footnote{Available from {\tt www.starlink.ac.uk}}
applications.  A bad pixel mask was created using \textsc{oracdr} and
chopped to 512 by 512 pixels.  The standard subtraction of dark frames
and flat fielding were carried out by \textsc{ccdpack}.  A 3D cube
consisting of the \textit{I, Q} and \textit{U} Stokes images, was
produced using \textsc{polka} from the \textsc{polpack} suite, and
this was then used to derive the per cent polarization, polarized flux
and polarization angle. A more detailed description of dual-beam
polarimetry and the data reduction techniques is given by \citet*{bg99}.

Photometric standards, FS 147 ({\it J}) and FS 141 ({\it K}), were
used to flux calibrate the data giving $J$=$11.18 \pm 0.04$ and
$K$=$10.29 \pm 0.12$.

For these observations, the focal plane polarimetry mask was removed,
so that a 512 by 512 pixel sub-array of the UIST detector could be
used.  This enabled faster read-out times and exposures of less than 1
second, so that observations of bright sources could be made without
the risk of saturation. This configuration of UIST resulted in the
overlapping of the \textit{o}- and \textit{e}-beams produced by the
Wollaston prism and a final analysis area of 20 by 60~arcsec. The
Wollaston prism splits each star into an \textit{e}- and
\textit{o}-component separated by 20 arcsec, so that any star in the
field lying more than 10 arcsec along the prism dispersion axis from
the target will only have one component in the analysis area.  Since
both \textit{e}- and \textit{o}-beams are required to correctly
calculate the Stokes intensities \textit{I}, \textit{Q} and
\textit{U}, these offset stars appear as highly polarized artefacts in
the reduced data, and they are marked as such on Fig. \ref{fig1}.
As the prism dispersion varies slightly with wavelength, this
results in an apparent shift of the artefact stars between the
\textit{J-} and \textit{K}-filters.

\label{jkpolres}

The \textit{J}- and \textit{K}-band polarimetric results are shown in
Fig.~\ref{fig1}. The total intensity images are shown in
Fig.~\ref{fig1} (a) and (c), superimposed with polarization
vectors, and show the centrally peaked nature of the source. The
object is clearly extended, relative to the 0.5~arcsec seeing FWHM,
with faint emission detected out to a radius of approximately
3~arcsec.  The lowest contour in both filters is 3 times the sky noise
and in the $I_{\rm J}$ image, shows that the faint emission is
elongated in a NNE/SSW direction. Details of contour levels are given
in the Figure caption. It is possible that a similar extension is
present in the $I_{\rm K}$ image, but confusion due to the presence
of the artefact stars makes this uncertain.



The polarized flux, produced by light scattering from dust grains, is
shown in Fig. \ref{fig1} (b) and (d). In both filters, the central
region appears elongated along a PA 136$\degr$ East of North, with two
bright shoulders of emission either side of the star. At {\it J}
(IP$_{\rm J}$ image) this structure is embedded within fainter more
extended emission orientated at 18$\degr$ East of North, seen in the
lowest three contours (the lowest contour is at 1.5 times the sky
noise). This faint extension is not as apparent in the $K$-band
polarized flux image (IP$_{\rm K}$), which is approximately
1~mag~arcsec$^{-2}$ shallower than the $J$-band data.  
The NW shoulder
is brighter than the SE shoulder, particularly apparent in the
IP$_{\rm J}$ image. Similar morphology has been observed in polarised
flux in a number of other PPN. \citet{gchy01} found bright arc-like
structures on either side of the star in {\it IRAS} 17436+5003 as well as
shoulder-like features in {\it IRAS} 19500-1709 and more ring-like
features in {\it IRAS} 22223+4327 and 22272+5435. 
They
interpreted these structures in terms of scattering from the inner
surfaces of a detached axisymmetric shell, with an equatorial density enhancement,
 and classified these objects
as ``shell-type'' objects. 
The arcs in {\it IRAS} 17436+5003 were
later fully resolved in mid-infrared imaging of thermal emission from
the dust \citep*{gy03} and successfully modelled using an axisymmetric
dust distribution based on that of \citet*{kw85}. 
Further evidence for
arcs and shoulders is seen in polarized flux images of {\it IRAS}
06530-0213, 07430+1115 and 19374+2359 \citep*{g05} and was interpreted
using light-scattering in a Kahn \& West density distribution.
We therefore interpret the polarized flux shoulders seen around
{\it IRAS} 19306+1407 in the same way, and suggest that they result 
from increased scattering at the inner boundary of a detached shell
with an equatorial dust density enhancement. 


The polarization vectors shown in Fig. 1 (a) and (c) are binned over
0.36 $\times$ 0.36 arcsec ($3\times 3$ pixels) and have a
signal-to-noise threshold of 2 in per cent polarization.  The vector
pattern appears approximately centro-symmetric in both filters,
indicating isotropic illumination by a central source.  The maximum
per cent polarization is $15 \pm 6$ and $10 \pm 4$ at \textit{J}- and
\textit{K}-bands respectively (Table~\ref{table2}).  These values are
lower limits to the intrinsic polarization, since in these
observations it has not been possible to correct for dilution of the
polarized flux by the unpolarized light from the central star.

\begin{table}
\caption{Summary of polarimetric results of \textit{IRAS}
19306+1407 for each band, detailing the maximum polarization,
integrated polarization and the position angle (E of N) of
the major and minor axis of the nebula in polarized flux. }
\label{table2}
\begin{tabular*}{\columnwidth}{ccccc}
\hline
 Band & Max. Pol.  & Integrated Pol.$^{\dagger}$ & PA$_{\rmn{major}}$ & PA$_{\rmn{minor}}$ \\
      & (per cent) & (per cent)      & ($\degr$)    & ($\degr$) \\
\hline
 {\it J}  & $15 \pm 6$ & $1.7\pm0.1$   & 18           & 136 \\
 {\it K}  & $10 \pm 4$ & $1.3\pm0.1$   & 18           & 136 \\
\hline
\end{tabular*}
$^{\dagger}$ - The integrated polarization over the source with
apertures of radii of 1.7- and 1.4-arcsec for {\it J-} and {\it K-}band respectively.
\end{table}

\subsection{\textit{Hubble Space Telescope} observations and results}
\label{hst}
We have obtained archived {\it HST} images for {\it IRAS}
19306+1407\footnote{Based on observations made with the NASA/ESA
Hubble Space Telescope, obtained from the data archive at the Space
Telescope Institute.  STScI is operated by the association of
Universities for Research in Astronomy, Inc. under the NASA contract
NAS 5-26555.}  observed on 2003 September 8 (proposal ID: 9463).  The
observations were obtained with the Advanced Camera for Surveys (ACS),
in conjunction with the High Resolution Channel (HRC), using {\it
F814W-} and {\it F606W}-filters with pivotal wavelengths of 5888 and
8115~\AA~respectively.  The images were reduced using the On-the-Fly
Reprocessing of {\it HST} Data ({\sc otfr}), which produces a
cosmic-ray cleaned, calibrated, geometrically corrected mosaic image.
Aperture photometry was performed using \textsc{gaia}, using the Vega
zero points\footnote{{\tt
http://www.stsci.edu/hst/acs/analysis/zeropoints}}, and obtained
magnitudes of $13.81 \pm 0.03$ and $12.45 \pm 0.02$ for {\it F606W} and {\it F814W}
respectively (Table \ref{table1}).

The reduced {\it F606W} and {\it F814W} images are shown in Fig. 2 (a)
and (b). Fig. 2 (c) shows the {\it F606W} image superimposed with
contours of {\it J}-band polarized flux from Fig. 1 (b). The object is
clearly bipolar in the {\it F606W} image, and the curved edges of
bipolar cavities, extending for 3 to 4 arcsec from the source, can be
seen. The orientation of the bipolar axis, at PA 18 deg, is aligned
with the {\it J}-band elongation in total and polarized intensity seen
in Fig. 1 (a) and (b). The bipolar structure appears to be surrounded
by a faint, more spherically symmetric halo, seen in both {\em HST}
filters, and this corresponds in extent to the outer contours in
Fig. 1 (a) and (c).  The polarized flux shoulders, at PA 136 deg, are
not perpendicular to the major axis of the nebula and this is clearly
seen in Fig. 2 (c). This non-orthogonality in the two axes will be
discussed further in Section 4.

The southern bipolar lobe appears to be the brighter of the two in both
{\em HST} filters, which could indicate that the major axis is slightly
inclined to the plane of the sky.

\begin{figure}
\begin{center}
\vspace{7.5cm}
\caption{
{\bf High resolution images are available at http://star-www.herts.ac.uk/$\sim$klowe/}.
The {\it HST} ACS images, scaled logarithmically, of {\it IRAS} 19306+1407.
(a) {\it F606W} (5888{\AA}) scaled between 22 and 13 mag~arcsec$^{-2}$ with
the angle of the major axis indicated by the arrow.
(b) {\it F814W} (8115{\AA}) scaled between 22 and 11 mag~arcsec$^{-2}$.
(c) {\it F606W} image, scaled as above, and {\it J-}band polarized flux contours.
The lowest contour level is 19 mag~arcsec$^{-2}$ and subsequent contours are
separated by 1 mag~arcsec$^{-2}$.}
\label{fig2}
\end{center}
\end{figure}

\subsection{Sub-millimetre observations and results}
Observations were made on 2005 January 8 using the Sub-millimetre
Common User Bolometer Array (SCUBA) at the 15~m James Clerk Maxwell
Telescope (JCMT) on Mauna Kea, Hawai'i.  The SCUBA observations were
made simultaneously at 450 and 850~$\umu$m in photometry mode using
a jiggle pattern.
The 450 and 850~$\umu$m photometry data were reduced
using the \textsc{surf} package within the \textsc{starlink} suite.
The sky opacity was corrected using the Caltech Sub-millimetre
Observatory (CSO) tau relationship\footnote{Using the revised 2000
October 25 relations}.  Flux calibration was performed using Mars,
inclusive of a maximum $\pm$5 per cent error due to the orientation of
Mars' poles relative to the Earth and Sun.  \textit{IRAS} 19306+1407
was detected at 450- and 850-$\umu$m at $>1\sigma$ and $>3\sigma$
respectively inclusive of calibration errors.  The fluxes obtained
(Table \ref{table1}) for $F_{450}$ and $F_{850}$ are $49.9 \pm 38.7$
mJy and $14.1 \pm 3.7$ mJy within a beam size of 7.5 and 14~arcsec
respectively.

\section{Modelling the CSE}
\label{mod:cse}
\subsection{Model details}
To investigate the dusty CSE around {\it IRAS} 19306+1407, we use
modified versions of the \citet{m89} axisymmetric light scattering
(ALS) code to produce Stokes {\it I, Q, U} images and the axisymmetric
radiative transfer (DART) code \citep{er90} to model the SED. Both
codes have previously been used to model the CSEs of post-AGB stars.
\citet*{gy03} used DART to simulate multi-wavelength mid-infrared
imaging observations of {\it IRAS} 17436+5003, in which an
axisymmetric shell was resolved. To simulate the axisymmetry, these
authors used a simple dust density formulation from \citet*{kw85}
which was found to successfully reproduce all of the axisymmetric
features, including the offset location of the brightness peaks seen
in the data, which was found to be due to the inclination of the
system to the plane of the sky. \citet*{g05} have used the ALS code to
produce generic light scattering models of PPN at varying optical
depth and also find that a Kahn \& West density model provides a good
representation of the observations with a minimum number of model
parameters. It is important that the dust density model uses a minimum
number of parameters whilst achieving an adaptable axisymmetric geometry, so
that there is a better chance of each parameter being observationally
well constrained. 
More complex dust density formulae have
been used \citep[e.g.][]{mubs02}, which incorporate the presence of AGB 
and superwind mass loss histories, but require more
parameters (twice as many in the case of \citealt{mubs02}).
These models result in morphologies that are
qualitatively similar to our simpler models, but are 
unlikely to be well constrained by our observations.
In both the ALS
and DART models we therefore use a simpler density profile from \citet{kw85}
to model an axisymmetric shell, whilst recognising its limited ability to
reproduce more complex morphologies:

\begin{equation}
\label{density1}
\rho(r,\theta) = \rho_0\left(\frac{r}{r_{\rmn{in}}}\right)^{-\beta}
\left(1 + \epsilon
\sin^{\gamma}\theta\right),
\end{equation}

where $\rho_0$ is the density at the pole ($\theta = 0\degr$) at the
inner radius, $r_{\rmn{in}}$, and $\beta$ specifies the radial density
distribution. The azimuthal density distribution is determined by
parameters $\epsilon$ and $\gamma$, which specify the equator-to-pole
density ratio ($1+\epsilon$) and the degree of equatorial enhancement,
respectively. An increase to $\gamma$ flattens the density
distribution, creating a more toroidal structure.

All parameters in Equation 1 are optimized in the model, apart from
$\beta$, which is fixed at a value of 2 due to a limitation of the
DART code, corresponding to constant mass-loss rate and expansion velocity
for the AGB wind.  The ALS density profile includes an extra
parameter, that restricts the axisymmetry to within a radius,
r$_{\rmn{{\sc sw}}}$, modifying Equation \ref{density1} to:

\begin{equation}
\rho(r,\theta)=\rho_{0}\left(\frac{r}{r_{\rmn{in}}}\right)^{-\beta}
\quad \rmn{when}\quad r>r_{\rmn{{\sc sw}}}.
\end{equation}
A power law size distribution is used with spherical grains of radius
$a$, between a minimum and maximum grain size of $a_{\rmn{min}}$
and $a_{\rmn{max}}$ respectively, and a power-law index, $q$,:

\begin{equation}
n(a) \propto a^{-q}\quad \rmn{for} \quad a_{\rmn{min}} \leq a \leq a_{\rmn{max}}.
\end{equation}

The inclination of the symmetry axis to the plane of the sky is not
known.  As mentioned in Section~\ref{hst}, the southern bipolar lobe
appears slightly brighter than the northern one in {\em HST} imaging
(Fig~2), which could indicate a small inclination to the plane of the
sky. Although the near-infrared images appear consistent with zero
inclination (e.g. they are similar to edge-on axisymmetric shell
models shown in \citealt{g05}), we consider the inclination angle to
be a free parameter and allow it to vary in steps of 10 deg.

The overall chemistry of the system is uncertain.  The results from
\citet{hvk00} suggest a C-rich nature based on emission features
consistent with C-rich PNe.  \citet{hkpw04} re-evaluated the
mid-infrared spectra and classified {\it IRAS} 19306+1407 as ``UIR
features coupled with emission from crystalline silicates'' suggesting
a dual chemistry nature. The dust species that have been considered in
our models are amorphous carbon (amC), silicon carbide (SiC) and
Ossenkopf cold silicates, and we have obtained the optical constants
from \citet{poyh93}, \citet{p88} and \citet{ohm92} respectively.


We ran a total of over 150 ALS and over 300 DART models to create a
model grid for the free physical parameters (Table~\ref{table3}). The
minimum and maximum grain sizes were investigated from 0.005 to
1~\micron, with a variable grain size spacing typically 0.005 to
0.02~\micron.  The grain size power law index was varied between 3.0
to 6.0 at increments of 0.5.  The radial density fall off exponent is
fixed at $\beta=2$ and cannot not be varied.  The bin widths for the
CSE parameters, common to both models are 1, 2, 10$\degr$ and
0.1$\times$10$^{-2}$ for the equator-to-pole contrast ($\epsilon$),
equatorial density enhancement ($\gamma$), inclination angle
($\theta$) and the ratio of the inner-to-outer radii ($r_{\rm
in}/r_{\rm out}$) respectively. The stellar temperature, $T_*$, was
investigated using a series of Kurucz models \footnote{{ \tt
http://kurucz.harvard.edu/grids.html}} with solar metalicities and
temperatures separated by 1000~K.


The ALS code is used to determine the best-fitting envelope parameters
based on the morphology, azimuthal profiles in polarized flux and
radial profiles of the percentage polarization and total intensities.
The ALS code is additionally used to
constrain the dust grain size by generating polarization information.
The ALS estimate of the grain size is an important input to the DART
calculations, which would otherwise suffer from a degeneracy between
grain size and outer CSE radius, both of which strongly influence the
long-wavelength tail of the SED.  The optical depths at 0.55, 1.2 and
2.2~$\umu$m are also derived from the ALS model and subsequently
inserted into the DART model.  The DART model fits to the SED are used
to constrain the temperature of the central star, inner-to-outer and
stellar-to-inner radii ratios.  The two codes were used to iteratively
produce a convergent model.

\begin{centering}
\begin{figure*}
\vspace{16cm}
\caption{{\bf High resolution images are available at http://star-www.herts.ac.uk/$\sim$klowe/}.
The 1.2- and 2.2-$\umu$m smoothed model images of
\textit{IRAS} 19306+1407 are displayed at the top and bottom of the
figure respectively.  These images are rotated to a PA of 136\degr to
mimic the observed data.  As with the observed images they have been
scaled logarithmically.  The total intensity (I) is displayed in
sub-figures (a) and (c) with overlaid polarization vectors (pol) and
polarized flux is shown in (b) and (d).  The model images have been
normalised at the same levels as the observed images: (a) and (c) are
scaled between 20 and 13 mag~arcsec$^{-2}$ with lowest outer contour
levels at 19 and 18 mag~arcsec$^{-2}$, respectively, separated by
1~mag~arcsec$^{-2}$; (b) and (d) are scaled between 20 to 16
mag~arcsec$^{-2}$ and 19 to 16 mag~arcsec$^{-2}$ respectively with
lowest outer contours at 19 (b) and 18 (d) mag~arcsec$^{-2}$,
separated by 0.5 mag~arcsec$^{-2}$.
}
\label{fig3}
\end{figure*}
\end{centering}

\begin{figure*}
\vspace{15cm}
\caption{{\bf High resolution images are available at http://star-www.herts.ac.uk/$\sim$klowe/}.
{\bf Left:} Azimuthally averaged radial profiles of the
normalised total intensity.  {\bf Centre:} Azimuthally averaged radial
profiles of the per cent polarization.  {\bf Right:} Radially averaged
azimuthal profiles of the normalised polarized intensity. In all
cases, the {\it J-} and {\it K-}band data are displayed as squares and
triangles respectively, with 3$\sigma$ error bars, and the 1.2- and
2.2-$\umu$m smoothed model data are displayed as solid and dashed
curves respectively. }
\label{fig4}
\end{figure*}

\begin{table}
\caption{The CSE and dust grain parameters
for the best-fitting ALS and DART models for \textit{IRAS} 19306+1407.}
\label{table3}
\label{table4}
\label{table7}
\begin{tabular*}{\columnwidth}{@{}lcl}
\hline
Parameter                  & Value     & Description                  \\
\hline
\multicolumn{3}{@{}l}{\textbf{Dust grain parameters}}                          \\
\multicolumn{3}{@{}l}{Ossenkopf} \\
Cold Silicates$^1$                       & 1.0 $\pm$ 0.01  & Number fraction               \\
$a_{\rmn{min}}$ ($\umu$m)                & 0.10 $\pm$ 0.01  & Minimum grain radius          \\
$a_{\rmn{max}}$ ($\umu$m)                & 0.40 $\pm$ 0.01  & Maximum grain radius          \\
$q$                                      & 3.5 $\pm$ 0.5   & Grain size power law index    \\
\noalign{\smallskip}
\multicolumn{3}{@{}l}{\textbf{Common envelope model parameters}} \\
$\beta^{\dagger}$                        & 2           & Radial density fall off \\
$\epsilon$                               & 6 $\pm$ 1   & Equator-to-pole density contrast \\
$\gamma$                                 & 5 $\pm$ 2   & Equatorial density enhancement  \\
$\theta$ (deg)                           & 0 $\pm$ 10  & Inclination angle (from equator) \\
$r_{\rmn{in}}/r_{\rmn{out}}$ (10$^{-2}$) & 7 $\pm$ 1 & Inner-to-outer radius ratio\\
\noalign{\smallskip}
\multicolumn{3}{@{}l}{\textbf{ALS model parameters}} \\
$\tau_{1.2}^{\ddagger}$ ($\times$10$^{-1}$) & 6.78 $\pm$ 0.05 & Optical depth at 1.2 $\umu$m   \\
$\tau_{2.2}^{\ddagger}$ ($\times$10$^{-1}$) & 1.13 $\pm$ 0.01 & Optical depth at 2.2 $\umu$m    \\
$r_{\rmn{{\sc sw}}}/r_{\rmn{in}}$         & 2.0  $\pm$ 0.5  & Super-wind \\
                                          &                 & to inner radius ratio\\
\noalign{\smallskip}
\multicolumn{3}{@{}l}{\textbf{DART model parameters}} \\
$T_{\star}$  (10$^3$~K)                  & 21   $\pm$ 1   & Effective Stellar Temperature \\
$r_{\star}/r_{\rmn{in}}$ (10$^{-5}$)     & 1.4  $\pm$ 0.2 & Stellar-to-inner radius ratio \\
$A^{\rmn{CSE}}_{V_{\mbox{ }}}$ (mag)                & 2.0  $\pm$ 0.1 & Equatorial optical extinction \\
\hline
\end{tabular*}
$^1$\citet{ohm92}. $^{\dagger}$This variable is fixed in our model code and cannot be varied.
$^{\ddagger}$The optical depth is an output of the ALS model.
\end{table}

\subsection{Model results}

\subsubsection{ALS model}

Before the raw model images can be compared with the polarimetric
observations, they must be smoothed to mimic the effect of the
atmosphere and telescope. We find that a simple Gaussian filter is
unable to reproduce the wings of the PSF effectively, which is
essential since the PSF wings have a critical effect on the percentage
polarization in the envelope where the intensity is low, at $r \ga
r_{\rmn{in}}$. To obtain a more realistic fit we use a Moffat filter
profile:

\begin{equation}
M(r) \propto \left[1 + \left(\frac{r}{\alpha_{\rmn{mof}}}\right)^2\right]
^{-\beta_{\rmn{mof}}},
\end{equation}
where $r$ is radius from the source and $\alpha_{\rmn{mof}}$ and
$\beta_{\rmn{mof}}$ are fitting parameters \citep*{mof69}.  The Moffat
parameters were calculated by fitting to the PSF of a bright field
star (Table~\ref{table5}) and their uncertainties were estimated by
examining the fit to the remaining field stars. The filter was then
applied to the raw ({\it I, Q } and {\it U}) model images, which were
then combined to obtain polarized flux and per cent polarization
values.

The resulting best-fitting smoothed model is shown in Fig.~\ref{fig3}
and the parameters used are displayed in Table~\ref{table4}. The model
reproduces the centrosymmetric polarization pattern and the observed
degrees of polarization in the {\it J}- and {\it K}-bands. The
polarized flux images show the shoulders seen in the observations, due
to the enhanced scattering at the inner edges of the axisymmetric
shell, where the dust density is greatest.  In Fig.~1, the observed
polarized flux images show a peak of emission at the location of the
star. Any mis-alignment of the bright, centrally-peaked images during
the data reduction stages will lead to a residual polarization at this
location.  Since the polarized flux peak is narrower than the seeing
disc size, we cannot treat it as significant. We do not see polarized
emission from the location of the star in the model images, since
forward-scattered light (i.e. scattering angles close to zero) is
strongly depolarized. Higher spatial resolution observations will be
required to investigate the polarization within 0.2 arcsec of the
star. If there is significant polarized emission from this region,
then an additional dust component would be required in the model.

The fit
was assessed by comparing the full grid of ALS models to the
polarimetric observations. In particular, the radial and azimuthal
profiles of the smoothed model images and the observations were
compared and the profiles for the best-fit model are shown in
Fig.~\ref{fig4}. The total intensity image radial profile fit
(Fig.~\ref{fig4} left) provides a check on the level of smoothing, and
shows an excellent fit to the observed intensity profile at both
wavelengths. The fit to the radial distribution of per cent
polarization (Fig.~\ref{fig4} centre) allows us to constrain the dust
grain parameters and optical depth. The maximum degree of polarization
produced by the model is very sensitive to the grain size distribution
so we consider that the grain size is well constrained. The radial
distribution of per cent polarization depends strongly on the optical
depth (and hence the dust density), since this determines the surface
brightness of the CSE relative to the unpolarized light from the
smoothed PSF. We determine an optical depth of 0.68 and 0.11 at $J$
and $K$ respectively, so that the CSE is optically thin in the
near-infrared. The axisymmetry parameters, $\epsilon$ and $\gamma$ are
determined by comparing azimuthal polarized flux profiles to the data
(Fig.~\ref{fig4} right). The best fit gives an equator-to-pole density
contrast of 7.




\begin{table}
\caption{The Moffat filter profile parameters, $\alpha_{\rmn{mof}}$ and
$\beta_{\rmn{mof}}$, for a bright field star at {\it J } \& {\it K}.}
\label{table5}
\begin{center}
\begin{tabular}{ccc}
\hline
Band    & $\alpha_{\rmn{mof}}$ & $\beta_{\rmn{mof}}$ \\
\hline
{\it J} & $3.95 \pm 0.06$           & $2.4 \pm 0.2 $ \\
{\it K} & $3.03 \pm 0.02$           & $2.2 \pm 0.3 $ \\
\hline
\end{tabular}
\end{center}
\end{table}

\subsubsection{DART model}
\label{mod:DARTmodel}
\begin{table}
\caption{Photometric values for {\it IRAS}
19306+1407 collated from the literature:
(1)  \citet{hvk00}; (2) \citet{metal03};
(3) {\it MSX} Bands \citep{epkmcwecg03}, and
(4) \citet{iras88}.}
\label{table6}
\begin{center}
\begin{tabular*}{\columnwidth}{cccc}
\hline
Band  & Central         & Flux density & Reference \\
      & wavelength      &              & \\
      & ($\umu$m)       & (Jy)         & \\
\hline
{\it V} & 0.55                 & $ 7.40\times10^{-3} $ & (1) \\
{\it R} & 0.44                 & $ 2.21\times10^{-2}$ & (2) \\
{\it MSX A} & 8.28              & 1.16                  & (3) \\
{\it IRAS} 12$\umu$m & 12.0     & 3.58                  & (4)\\
{\it MSX C} & 12.13             & 3.65 & (3) \\
{\it MSX D} & 14.65             & 9.12 & (3) \\
{\it MSX E} & 21.34             & 46.27 & (3) \\
{\it IRAS} 25$\umu$m            & 25.0     & 58.65 & (4) \\
{\it IRAS} 60$\umu$m            & 60.0    & 31.83 & (4) \\
{\it IRAS} 100$\umu$m           & 100.0    & 10.03 & (4) \\
\hline
\end{tabular*}
\end{center}
\end{table}

\begin{figure}
\vspace{8cm}
\caption{
{\bf High resolution images are available at http://star-www.herts.ac.uk/$\sim$klowe/}.
The observed SED and best model fits for
\textit{IRAS} 19306+1407.
The dash line is the model fit and the
solid black line is the model fit with
interstellar reddening applied.
 References: (1) this paper, (2) \citet{hvk00},
 (3) \citet{ummpd03}, (4) \citet{metal03},
(5) \citet{epkmcwecg03} and (6) \citet{iras88}.}
\label{fig5}
\end{figure}

The SED of \textit{IRAS} 19306+1407 is plotted in Fig.~\ref{fig5}
using published photometry and spectroscopy from a variety of sources,
including this paper, and covering wavelengths from the $V$-band
through to the sub-millimetre. The photometric values are listed in
Table \ref{table6}. The double-peaked nature of the SED is immediately
evident, consisting of a reddened stellar peak around $1.6$~$\umu$m
and a broad thermal dust peak between 30 and 40~$\umu$m due to the
CSE.  Double-peaked SEDs are typical of post-AGB stars with optically
thin detached CSEs \citep*{vdvhg89}.

Our best-fitting model is shown in Fig.~5, both with and without
correction for interstellar extinction (see below). Previous
attempts to model the SED using amorphous carbon dust and a cooler F/G
type star, were found not to provide sufficient flux in the dust peak
\citep{hvk00}. We have treated the stellar temperature as a free
parameter and determined a best-fitting value of 21,000 K, typical of
a B1 type star. This is consistent within errors with the
observationally determined spectral type of B0: \citep{vhk04,kh05},
where the colon denotes an uncertainty in the 0
(Hrivnak, private communication).

An optical extinction of $A_{\rmn{V}} = 2.0$~$\pm$~0.1~mag, through
the CSE in the equatorial direction, was determined from the model
fit.  We have investigated the effect of inclination of the
nebula axis and have determined that the SED
is consistent with a value of $0\degr \pm 10\degr$. The extinction through
the CSE along our line of sight is, therefore, also $A_{\rmn{V}} = 2.0$.

{\it IRAS} 19306+1407 lies close to the galactic plane, $l=
50.30\degr$ and $b= -2.48\degr$, and the SED will be affected by
interstellar extinction.  The extinction through the Galaxy at this
point is estimated to be $A_V =$~5.1~$\pm$~0.2~mag.  This value was
obtained from the {\it IRAS} dust reddening and extinction
service\footnote{{\tt
http://irsa.ipac.caltech.edu/applications/DUST}}, based on the data
and technique in \citet*{sfd98}.

To correct the emergent model flux for interstellar extinction, we
apply a reddening model developed by \citet*{ccm89} which gives the
extinction, $A_{\lambda}$, at every wavelength between 0.1 and
3.3~$\umu$m for a given $A_{\rmn{V}}$ and extinction ratio,
$R_{\rmn{V}}$.  The extinction at shorter and longer wavelengths has
been extrapolated.  The DART model flux, $F_{\rmn{DART}}$, is then
modified to give the flux after correction for interstellar
extinction, $F_{\lambda}$:
\begin{equation}
\label{sedism}
F_{\lambda} = F_{\rmn{DART}} \times 10^{-\frac{A_{\lambda}}{2.5}},
\end{equation}

Assuming a standard value of $R_V$=3.1 for the ISM, then a fit to the
SED shortward of 6~$\umu$m gives a value of 4.2~$\pm$~0.1~mag for
interstellar extinction (solid curve in Fig.~\ref{fig5}). The total
extinction to the star is, therefore, 6.2~$\pm$~0.2~mag.  This is
consistent with the observed $J$-$K$ colours. Assuming an extinction
ratio ($R_V$) of 3.1, and an intrinsic colour excess of $E(J-K)_0 =
-0.09$ for a \textsc{B1I} star, gives $A_{\rm V}=6.4 \pm 0.7$~mag.
The model parameters used in DART are presented in Table~\ref{table7}.



\subsubsection{Distance estimate and derived parameters}

The interstellar extinction can be used to estimate the distance of
the post-AGB star.  Using \citet{j05}, based on extinction towards
open clusters, gives an estimated extinction of $1.58 \pm 0.04$~mag
kpc$^{-1}$.  A visual extinction of 4.2~$\pm$~0.1~mag suggests a
distance of $2.7 \pm 0.1$~kpc, which we now adopt as our assumed
distance from this point onwards.

Using this distance estimate gives values for $r_{\rmn{in}}$ and
$r_{\rmn{out}}$ of $1.9 \pm 0.1 \times$10$^{14}$ and $2.7 \pm 0.1
\times$10$^{15}$~m respectively.  Multiplying $r_{\rmn{in}}$ by
$r_{\star}/r_{\rmn{in}}$ gives a stellar radius, $R_{\star}$, of $3.8
\pm 0.6$~R$_{\sun}$.

The stellar luminosity, $L_{\star}$, is obtained by calculating the
integrated flux under the model SED, giving values of $1800 \pm 140$
and $4500 \pm 340 $~L$_{\sun}$, with and without interstellar
reddening applied respectively, for the assumed distance.  Post-AGB
stellar evolution models suggest a lower limit of 2500~L$_{\sun}$ for
the central star of a PN \citep{s83}, which means that {\it IRAS}
19306+1407 must be at least 2.0~kpc away to satisfy this criterion.

To calculate the time scales of mass loss, $r_{\rmn{in}}$ and
$r_{\rmn{out}}$ are divided by the AGB wind speed.  Only the H$_2$ and
H$\alpha$ kinematic information are available for {\it IRAS}
19306+1407.  These speeds arise from the shocks and fast winds in the
post-AGB phase, and are not a true reflection of the AGB envelope
expansion speed, therefore we have assumed a typical speed of
15~km~s$^{-1}$ from \citet{nklbl98}.  The age of the CSE is then $5700
\pm 160 $~yrs, became detached $400 \pm 10$~yrs and the mass loss
lasted $5300 \pm 160 $~yrs.

The number density of dust grains, $N_0$, at $r_{\rmn{in}}$ is
calculated from the optical depth, the extinction cross section of the
dust and the CSE thickness. The optical depth at 1.2~$\umu$m is $0.678
\pm 0.005$, giving a value of $N_0$ = $6.1 \pm 3.0
\times$10$^{-3}$~m$^{-3}$.  Using $N_0$ and integrating the dust
density distribution gives the total dust mass ($M_{\rmn{d}}$), and
assuming a dust grain bulk density of 3$\times$10$^3$~kg~m$^{-3}$,
gives a value of $8.9 \pm 5.0 \times$10$^{-4}$~M$_{\sun}$.

The gas-to-dust ratio for this object is unknown and we have adopted a
value of 200 from \citet{hh05}.  The total mass of the CSE is then
$1.8 \pm 1.0 \times$10$^{-1}$~M$_{\sun}$ with an average mass-loss
rate ($\dot{M}$) of $3.4 \pm 2.1
\times$10$^{-5}$~M$_{\sun}$~yr$^{-1}$.
The derived parameters given in this section are summarized in
Table~\ref{table8}.

\section{Discussion}

\subsection{CSE geometry}

The polarimetric observations, shown in Fig.~1, have been interpreted
in terms of an axisymmetric shell with an equatorial density enhancement, 
which is optically thin in the
near-infrared. The shell model successfully reproduces the observed
SED from the V band to the sub-millimetre. As a further check on the
validity of the model, the ALS code was run at the central wavelength
of the {\it F606W} filter to simulate the {\it HST} observations shown
in Fig.~2 (a). The results are shown in Fig.~6 and we find that the
bipolar structure is reproduced, inclusive of the flattened contours
in the centre of the {\it HST} image. A single axisymmetric shell
model, based on the simple \citet*{kw85} density distribution, can
account for the morphology of this object over a wide range of
wavelengths. The transition from bipolar nebula in the optical to
limb-brightened shell in the near-IR is due to the variation in
optical depth through the envelope with wavelength. At the wavelength
of the {\em HST} observations, the CSE is optically thick along the
equatorial direction and so light is preferentially funnelled along
the polar axes before scattering into our line of sight, creating the
bipolar lobes. The fact that the general appearance and extent of the
lobes is reproduced by the model indicates that the density structure
of the shell, in particular the equator-to-pole density contrast of 7,
is reasonable. At near-infrared wavelengths, where the shell is
optically thin along the equator, light is mainly scattered at the
inner boundary in the equatorial plane, where the dust density is
greatest, creating the shoulders seen in polarized flux in our
observations.

Since our model calculations are limited to axisymmetric geometries,
one aspect of the observations that we have not been able to account
for is the non-orthogonality of the polarized flux shoulders, at PA
136~deg, and the major axis of the nebula, at PA 18~deg, illustrated
in Fig.~2.  A similar `twist' has been detected in the mid-infrared
images of {\it IRAS} 17456+5003, which has a curving polar axis
\citep*{gy03}, and which was also modelled with an axisymmetric dust
shell. A further similarity between the two objects is the unequal
brightness of the polarized flux shoulders \citep[see][]{gchy01}. In
the context of our model, these are due to scattering at the inner
edge of the axisymmetric shell, so that the scattering optical depth
is greater on one side of the shell than the other. Assuming that the
dust properties are the same throughout the shell, then this suggests
that there is a greater concentration of dust in the brighter
shoulder. Further evidence for asymmetric dust distributions around
post-AGB stars is seen in mid-infrared images of {\it IRAS} 07134+1005
\citep{d98} and {\it IRAS} 21282+5050 \citep{m93}. \citet*{gy03}
discuss possible causes for these asymmetries and conclude that they
may arise due to interaction of the mass-losing star with a binary
companion, although exactly how this happens is not clear.

\citet{vhk04} imaged {\it IRAS} 19306+1407 using a narrowband H$_2$
filter (2.12~$\umu$m) and a narrowband {\it K} continuum filter
(2.26~$\umu$m), to investigate the molecular hydrogen emission.  Their
continuum subtracted H$_2$ image (their Fig.~2 left) shows a broken
ring with limb-brightened edges, which appears cospatial with the
central dust structure, at PA 136, seen in our polarized flux images.
The ring can also be seen in their 2.26~$\umu$m continuum image, so
that they have resolved the dust structure that we see in polarized
flux. The similarity between the polarized flux and H$_2$ images
suggests that the scattered light and molecular emission originate in
the same region.  \citet{vhk04} also detect faint extended H$_2$
emission lobes, extending from the ring, corresponding to the extended
bipolar structure seen in the HST images (Fig.~\ref{fig2}), oriented
PA 18$\degr$. It appears that the same axis twist seen in the
scattered light images may be present in H$_2$ emission. \citet{vhk04}
suggest that the H$_2$ ring seen in their images collimates the
H$_2$-emitting bipolar lobes.



\subsection{Estimation of the dust mass from our sub-mm observations}
\label{estdustsub}
The mass of dust in the CSE, $M_{\rmn{d}}$, can be estimated from the
{\it IRAS} 100~$\umu$m flux, $F_{100}$, and the SCUBA 850~$\umu$m
flux.  We have used the method stated in \citet*{gby02} to calculate
an estimate of the dust mass from our observations.  The dust
temperature is estimated to be 146~$\pm$~21~K, using Wien's
displacement law, with the peak dust emission at $35 \pm 5$~$\umu$m.
The 850~$\umu$m flux value given in Table~\ref{table1} and $F_{100}=
10.03 \pm 1.30$~Jy, gives an emissivity index of 1.3~$\pm$~0.1.  The
assumed density for a silicate dust grain is
3$\times$10$^3$~kg~$\rmn{m}^{-3}$.  The total dust mass in the CSE,
using the assumed distance, is then $4.3 \pm 0.7
\times$10$^{-4}$~$\rmn{M_{\sun}}$, which is a factor of $\sim 2$ less
than the value obtained from our radiative transfer model.  The
difference may arise from the simple assumptions inherent in the
sub-millimetre estimate, particularly that of an isothermal CSE. The
bulk of dust in the envelope will be cooler than 146~K (the maximum
and minimum dust temperatures in the DART model are 130 and 40~K
respectively), and will radiate on the long wavelength tail of the
SED. An isothermal temperature of 100~K would result in a dust mass of
$7.2 \pm 1.7 \times$10$^{-4}$~$\rmn{M_{\sun}}$. Given these
approximations, we consider that the two results are comparable but
that the more rigorous model calculations from DART and ALS provide a
realistic value for the dust mass in the CSE.

\subsection{CSE chemistry}
\label{dis:geochem}

\begin{centering}
\begin{figure}
\vspace{8cm}
\caption{{\bf High resolution images are available at http://star-www.herts.ac.uk/$\sim$klowe/}.
A comparison of the {\it F606W} {\em HST} image and the raw
model image from ALS at the central wavelength of the {\it F606W}
filter. The model image has been rotated parallel to the long axis (PA
= 18$\degr$) to match the {\it HST} image.  The contours are spaced at
an interval of 1~mag~arcsec$^{-2}$ from the peak value.}
\label{fig6}
\end{figure}
\end{centering}


We have modelled {\em IRAS} 19306+1407 using a silicate dust model,
with grain sizes between 0.1 and 0.4~$\umu$m, which reproduces the
shell-like morphology in the near-infrared, the observed degrees of
polarization and the SED. However, we also find that a purely C-rich
chemistry (amorphous carbon) using larger grains, typically
$>$0.6~$\umu$m, can reproduce the observed polarization \citep{lg05}
and fit the overall shape of the SED, although this produces a poor
fit at $<$1~$\umu$m after interstellar reddening is applied. Amorphous
carbon also does not reproduce the shape of the SED between 10 and
20~$\umu$m. We have investigated the possibility that silicon carbide
could fit the 10-20~$\umu$m region, but find that it provides too much
flux at 11-12~$\umu$m and was in general a poor fit to the SED.  These
regions are modelled more effectively using Ossenkopf cold silicates.



As mentioned in Section~3.1, the simultaneous presence of emission
from PAHs and crystalline silicates \citep{hvk00,hkpw04} suggests that
the CSE has a mixed chemistry (both O- and C-rich). Our simple
investigations of mixes of carbon and silicate dust in the CSE, show
that amorphous carbon significantly dominates the SED at less than 1
per cent abundance. This suggests that if the 10-20~$\umu$m fits
require silicate grains, then they must be the dominant dust
component. However our models do not allow us to segregate the O- and
C-rich material to have, for example, a region of silicate grains
close to the star with a largely C-rich outflow at larger radii. Such
a configuration has been proposed to explain observations of mixed
chemistry objects \citep{mol02} in which the crystalline emission
comes from cool silicates trapped in stable circumstellar or
circumbinary discs.  \citet{mat04} have shown that in the mixed
chemistry post-AGB object {\em IRAS} 16279-4757 the carbon-rich dust,
traced by PAH emission, is located in a low-density outflow, while the
continuum emission is concentrated toward the centre. Although our
single component model, based on silicate grains, is reasonably
successful in reproducing the observations, it is almost certain
that the chemistry of {\em IRAS} 19307+1407 involves both O- and
C-rich material, perhaps spatially segregated and with more than one
size distribution.


\begin{table}
\caption{The derived model parameters at the assumed distance of
2.7~kpc obtained from ALS$^{\dagger}$ and DART$^{\ddagger}$ models.}
\label{table8}
\begin{tabular*}{\columnwidth}{l@{\extracolsep{\fill}}rll}
\hline
Parameter               & Value           & Units                & Description  \\
\hline
$R_{\star}$            & 3.8  $\pm$ 0.6 & R$_{\sun}$            & Stellar Radius$^{\ddagger}$ \\
$r_{\rmn{in}}$         & 1.9  $\pm$ 0.1 & (10$^{14}$) m         & Inner Radius$^{\dagger \ddagger}$ \\
$r_{\rmn{{\sc sw}}}$   & 3.8  $\pm$ 1.0 & (10$^{14}$) m         & Super-wind Radius$^{\dagger}$\\
$r_{\rmn{out}}$        & 2.7  $\pm$ 0.1 & (10$^{15}$) m         & Outer Radius$^{\dagger \ddagger}$ \\
$L^{\diamond}_{\star}$ & 4500 $\pm$ 340 & L$_{\sun}$            & Stellar Luminosity$^{\ddagger}$ \\
$N_0$                  & 6.1  $\pm$ 3.0 & (10$^{-3}$) m$^{-3}$  & Number density at $r_{\rmn{in}}^{\dagger}$ \\
$M_{\rmn{d}}$          & 8.9 $\pm$ 5.0  & (10$^{-4}$) M$_{\sun}$ & Total mass of Dust$^{\dagger}$ \\
$A_{\rmn{V}}$          & 4.2 $\pm$ 0.1  & mag                   & Interstellar extinction$^{\ddagger}$ \\
$T_{\rmn{max}}$        & 130 $\pm$ 30   &  K                    & Temperature at $r_{\rmn{in}}^{\ddagger}$ \\
$T_{\rmn{min}}$        & 40  $\pm$ 20   &  K                    & Temperature at $r_{\rmn{out}}^{\ddagger}$ \\
\hline
\end{tabular*}
$^{\diamond}$The apparent luminosity of the star,
with applied interstellar reddening, is 1800~$\pm$~140~L$_{\sun}$.
\end{table}

\section{Conclusion}
We present near-infrared polarimetric images of the dusty CSE of {\it
IRAS} 19306+1407, in conjunction with new submillimetre photometry and
archived {\it HST} images.  The polarization vectors show a
centrosymmetric structure with a maximum polarization of $15 \pm 6$
and $10 \pm 4$ per cent for {\it J-} and {\it K-}band respectively.
The polarized flux shows a very faint elongated distribution at PA
18$\degr$ with two bright scattering shoulders at PA 136$\degr$. The
object is clearly bipolar in archived {\em HST} images, with the bipolar
axis also at PA 18$\degr$

We model the polarimetric data using an axisymmetric light scattering
code and a dust model based on sub-micron sized silicate grains. The
observed polarization features are well described by a simple
axisymmetric shell geometry, with an equator-to-pole density contrast
of 7.  The same shell model is used to fit the SED of {\it IRAS}
19306+1407 from optical to sub-millimetre wavelengths using an
axisymmetric radiation transport code, to constrain the stellar
temperature and radius, the optical depth of the CSE and the mass of
dust in the CSE. We find that a B-type stellar spectrum, with
T$_*$=21,000~K, best describes the SED, confirming previous
suggestions that the object is an early PN.

The models give a value for the CSE and interstellar extinction of
2.0~$\pm$~0.1~mag and 4.2~$\pm$~0.1~mag respectively.  We estimate a
distance, from the interstellar extinction, of 2.7~$\pm$~0.1~kpc and
use this value to derive parameters from our models.

The polarimetric imaging shows deviations from axisymmetry that are
beyond the scope of our model calculations. There appears to be a
greater concentration of dust on one side of the star than the other,
plus the axisymmetric shell is not aligned with the larger-scale
bipolar axis, clearly seen in archive {\em HST} images. Similar
features are seen in other post-AGB objects and may result from
interaction of the mass-losing star with a binary companion.  Further
evidence for a binary nature is provided by the probable mixed
chemistry nature of this object.

\section{Acknowledgements}
Krispian Lowe
is supported by a PPARC studentship.
The United Kingdom Infrared Telescope is operated by the
Joint Astronomy Centre on behalf
 of the U.K. Particle Physics and Astronomy Council (PPARC).
The James Clerk Maxwell Telescope is operated by
The Joint Astronomy Centre on behalf of the
Particle Physics and Astronomy Research Council
of the United Kingdom, the Netherlands Organisation
for Scientific Research, and the National Research Council of Canada
(Program ID: S04BU09).
Based on observations made with the NASA/ESA Hubble Space Telescope,
obtained from the data archive at the Space Telescope Institute.
STScI is operated by the association of Universities for
Research in Astronomy, Inc. under the NASA contract  NAS 5-26555.
All model calculations were run on the HiPerSpace Computing
Facility at University College London.
Thank you to Kim Clube for discussions on DART modelling.
A thank you to Bruce Hrivnak for the private communication
on \textit{IRAS} 19306+1407.
VizieR catalogue access tool and {\sc SIMBAD} database, {\sc CDS},
Strasbourg, France.
This research has made use of NASA's Astrophysics Data System.

\end{document}